\documentclass[sn-nature]{sn-jnl}

\usepackage{graphicx}%
\usepackage{multirow}%
\usepackage{amsmath,amssymb,amsfonts}%
\usepackage{amsthm}%
\usepackage{mathrsfs}%
\usepackage{hyperref}
\usepackage[title]{appendix}%
\usepackage{xcolor}%
\usepackage{textcomp}%
\usepackage{manyfoot}%
\usepackage{booktabs}%
\usepackage{algorithm}%
\usepackage{algorithmicx}%
\usepackage{algpseudocode}%
\usepackage{listings}%
\usepackage{pdfpages}
\usepackage[export]{adjustbox}
\makeatletter
\renewcommand\@biblabel[1]{#1.}
\makeatother

\begin{document}

\title[Article Title]{Intermittent turbulent gusts lift eagles }

\author*[1]{\fnm{Dipendra} \sur{Gupta}}\email{dg535@cornell.edu}
\author[2]{\fnm{David} \sur{Brandes}}
\author[3]{\fnm{Michael J} \sur{Lanzone}}
\author[4]{\fnm{Tricia} \sur{Miller}}
\author*[1]{\fnm{Gregory P} \sur{Bewley}}\email{gpb1@cornell.edu}

\affil[1]{\orgdiv{Sibley School Mechanical and Aerospace Engineering}, \orgname{Cornell University}, \orgaddress{\city{Ithaca}, \postcode{14850}, \state{NY}, \country{USA}}}

\affil[2]{\orgdiv{Civil \& Environmental Engineering}, \orgname{Lafayette College}, \orgaddress{\city{Easton}, \postcode{18042}, \state{PA}, \country{USA}}}

\affil[3]{\orgname{Cellular Tracking Technologies}, \orgaddress{\city{Rio Grande}, \postcode{08242}, \state{NJ}, \country{USA}}}

\affil[4]{\orgdiv{Conservation Science Global, Inc.}, \orgaddress{\city{Cape May}, \postcode{08204}, \state{NJ}, \country{USA}}}

\abstract{Turbulence grounds aircraft 
and combating it in flight requires energy, 
yet volant wildlife fly effortlessly even on windy days. 
The nature of the interactions between soaring birds 
and transient turbulent gusts is not clear, 
especially when compared with our understanding of flight in larger and steadier airflows during thermal or dynamic soaring. 
We show that soaring golden eagles ({\it Aquila chrysaetos}) experienced short upward accelerations indicative of preferential engagement with strong and intermittent turbulent updrafts. 
The vertical accelerations reflect changes in lift that were as large as 25 standard deviations from the mean, 
or more than three times the acceleration of gravity, 
and so large as not to be consistent with gust mitigation or avoidance. 
These extreme events occurred in short bursts that mimic movement {\it with} turbulent vortices. 
The burst statistics and their symmetries approach those of turbulence 
toward longer timescales. 
On the shortest timescales, 
the bursts break the symmetry of small-scale turbulence 
in favor of upward accelerations that are more intermittent than turbulence. 
We introduce a simple nonlinear model that predicts the scale at which symmetry breaks and the stronger intermittency on the smaller scales. 
These findings suggest a ratcheting mechanism on turbulent gusts and constitute the first quantitative evidence in favor of turbulent gust harvesting by wildlife. 
An implication is that turbulence is so strong and pervasive as to make unsteady and nonlinear aerodynamics an intrinsic and beneficial aspect of both flapping and soaring flight in the atmospheric boundary layer -- 
one that we need to incorporate in our understanding of the energetics of flight.  }


\maketitle


Understanding the relationships between environmental turbulence, gusts, and flight 
is more important than ever due to the advent of small semi-autonomous 
as well as slow-moving aerial vehicles 
that are more susceptible to turbulence 
than their larger, heavier, and faster counterparts. 
The effort to understand how to fly in turbulence complements our broader work to understand the relationships between wildlife and movements of the atmosphere, 
which is driven in part by an increasing awareness of the impact of climate change and human development on wildlife, 
\cite{newton1998population, jonzen2006rapid, nourani2017effects, hejazi2023honeybee}
and the opportunity to use wildlife motions to monitor the state of the atmosphere. 
\cite{laurent2021turbulence, lempidakis2022estimating}

Turbulence is multiscale -- 
it is composed of superimposed vortical motions on a continuous distribution of length and time scales. 
In the atmosphere, the timescales range from fractions of a second to hundreds of seconds. \cite{boettcher2003statistics}
Animal behavior, on the other hand, tends to concentrate on particular scales. 
Flapping, for instance, occurs at a frequency of about 2.5\,Hz in golden eagles. \cite{laurent2021turbulence}
Turbulence and flapping are both intermittent. 
Golden eagles switch between flapping, gliding, and other flight modes 
in ways that are correlated with atmospheric conditions. 
\cite{lanzone2012flight,duerr2012testing,bohrer2012estimating}
Turbulence is intermittent in the sense that periods of intense activity are interspersed with periods of relative calm. 
\cite{batchelor1949nature}
While turbulence is more intricate than the black-and-white nature of switching between flight modes, 
the mark of turbulent intermittency 
is that strong fluctuations, or gusts, 
occur many orders of magnitude more often than Gaussian models predict, 
especially in the atmosphere. 
\cite{bottcher2007small}
A defining feature of turbulence is that intermittency is more pronounced toward smaller and shorter scales in specific and quantified ways. 
\cite{sreenivasan1997phenomenology}

In engineering practice, 
turbulence has usually been considered a disturbance, 
and one that needs to be mitigated or avoided. 
Flight vehicles are designed to operate efficiently under calm conditions with corrections added to account for turbulence and gusts. 
\cite{mohamed2014attitude, jones2022physics} 
Natural fliers, on the other hand, 
seem to be capable of flying effortlessly, even gracefully, 
in extremely turbulent conditions; 
as if they are surfing on the turbulence. 
This raises the fundamental question of whether there are conditions under which we may not need to suppress, mitigate, reject or avoid turbulence and gusts, 
and conditions under which we can exploit turbulence. 
To find out, it is useful, if not crucial, to understand the influence of turbulence intermittency on volant wildlife. 

Large birds soar, or fly without flapping, when they migrate over long distances or search for prey. 
To reduce their energy expenditure, they exploit naturally available energy within the boundary layer of the atmosphere. 
Energy is available in three forms in principle: 
(i) ascending vertical air currents such as those in thermal updrafts, 
\cite{pennycuick1998field, reddy2016learning, akos2010thermal, kerlinger1989flight}
in orographic updrafts, \cite{kerlinger1989flight,bohrer2012estimating}
or in mountain waves, \cite{bohrer2012estimating}
(ii) mean wind gradients such as those harnessed during dynamic soaring, 
\cite{rayleigh1883soaring, richardson2015upwind}
and (iii) fluctuating gradients in turbulence and gusts.  
\cite{patel2009extracting, mallon2016flight, laurent2021turbulence, bollt2021extract, he2023energy}
The spatiotemporal scales at which these sources operate are different. 
Thermal updrafts and wind shear, for instance, are relatively stable large-scale structures that evolve slowly relative to the time it takes to fly through them, 
and are predictable enough to have been exploited by glider pilots, birds, and autonomous flight vehicles. 
\cite{akos2010thermal, richardson2015upwind, reddy2018glider}
Small-scale gusts and turbulence, on the other hand, are largely stochastic, \cite{bandak2024spontaneous}
and so far have generally been regarded by humans as being responsible for energetic costs, 
\cite{reynolds2014wing, lempidakis2024turbulence}
rather than energetic reservoirs. 
\cite{langley1893internal, mallon2016flight, bollt2021extract}
Despite its stochastic nature, we know from computer simulations that turbulence can in principle be exploited to achieve energy-efficient flight even when only local information is available about the flow, and we know that these strategies are associated with increased accelerations. 
\cite{bollt2021extract}
Strategies to harness vortical flows have recently received computational attention for underwater navigation. 
\cite{biferale2019zermelo, gunnarson2021learning, monthiller2022surfing, xu2023long}
Quantifying the energetics of animals in practice, however, is difficult. 
\cite{engel2005racing, schmidt2007energy, friman2024pays}

Turbulence appears to be stronger than any other contribution 
to the accelerations of soaring golden eagles 
on timescales between about 0.1 and 10\,s. 
\cite{laurent2021turbulence}
In this interval, 
the response of the eagles to typical turbulent fluctuations can be explained with a linear model. 
Typical fluctuations in atmosphere are characterized by 
velocities on the order of 1\,m/s. 
In this paper, 
we examine the response of eagles to extreme fluctuations, 
those characterized by much larger velocities, 
and we ask whether the approximate linearity persists. 
It appears not to persist in an interesting and nontrivial way. 

The relevance of intermittent gusts, 
or of extreme wind velocity fluctuations, 
to the aerodynamics of flight is still not clear. 
Gusts occur on timescales as short as fractions of a second 
and can generate velocities on the order of 10\,m/s, 
large enough in principle to cause non-linear aerodynamic responses. 
\cite{mohamed2023gusts}
In steady state, aerodynamic stall causes the lift force to decrease if the angle of attack of a wing exceeds approximately 15 degrees 
(Fig.~1, \textit{inset}), \cite{anderson2011ebook}
a response that a sustained gust can elicit. 
\cite{biler2021experimental, fukami2023grasping}
Eagles might take deliberate and nonlinear actions in response to gusts either to mitigate them, 
\cite{reynolds2014wing}
or potentially to exploit them and to increase the lift force transiently. 
Active responses, however, are not needed to generate transient lift increases. 
They have been observed to arise 
from passive unsteady aerodynamic effects 
such as the development of leading edge vortices. 
\cite{eldredge2019leading}

\subsection*{How we related turbulence, gusts and eagle accelerations}
In order to analyze the effects of gusts, we may model 
the temporal profile of the gust wind velocity 
using cosine, linear ramp, or boxcar shapes. 
\cite{jones2022physics}
Unlike turbulence, which is multi-scale, 
models like these typically embody only a single timescale and amplitude, 
which limits their ability to capture realistically the motions in the atmosphere. 
To incorporate the stochastic nature of real gusts, 
power spectral and coherence functions have been introduced, 
\cite{solari1987turbulence, bierbooms2002stochastic}
but these assume that gusts are generated by Gaussian random processes, 
which they are not. 
As a consequence, the extreme fluctuations generated by turbulence at small scales are ignored. 
\cite{jones2004documentation, liu2010probability}

To identify turbulent gust responses in eagles, 
we use statistical tools established for classical turbulence \cite{sreenivasan1997phenomenology}
including the conditionally averaged velocity difference profiles studied by Mouri {\it et al.}~\cite{mouri2003vortex}
This latter metric has the advantage of embodying the notion of continuous distributions of two types: 
of amplitudes and of scales, 
the scales being the distances between the two points composing the velocity difference. 
The statistic is empirical, and so reflects the structure of real turbulent gusts. 
Experiments show that the conditional statistics resemble the cross-section of 
idealized Burgers vortices, 
and that they depend only weakly on scale or amplitude. 
\cite{burgers1948mathematical, mouri2003vortex}
In this picture, each vortex forms both an updraft and a downdraft whose symmetry increases toward smaller and shorter scales. 
\cite{kolmogorov1991local, kurien2000anisotropic, biferale2005anisotropy}
The signatures of turbulence and intermittency in this statistic 
as well as in the velocity difference flatness described below are known quantitatively. 
\cite{sreenivasan1997phenomenology}

We analyzed the accelerations measured by instruments attached to the backs 
of freely flying wild golden and bald eagles. 
\cite{miller2019implications, garstang2022instrumented}
Since bald eagles flap more frequently ($\approx$38\% of total flight time) 
than golden eagles ($\approx$2\% of total flight time), 
we compared the two species to discriminate between 
the statistical signatures of flapping intermittency, 
which is concentrated on the timescale of flapping and dominates the bald eagle data, 
and turbulence intermittency, 
which is distributed across a broad range of timescales. 
The relative infrequency of flapping in golden eagles 
also made it possible to accumulate enough data 
for conditional statistics to converge. 


We examined statistics of the eagles' accelerations in the vertical direction, 
$a_z(t)$, 
and the acceleration differences, 
$\delta a_z(t, \tau) \equiv a_z(t+\tau/2) - a_z(t-\tau/2)$, 
where the differences $\delta a_z$ draw out activity in $a_z$ 
on the timescale of the time increment, $\tau$. 
\cite{kolmogorov1991local}
The acceleration differences embody contributions from interactions with turbulence 
and from all other elements of an eagle's airborne life. 
Among these elements, 
we distinguish only between gliding and flapping flight, 
and to do so we take advantage of the fact that flapping in eagles is essentially a binary phenomenon: 
the eagles were either flapping or not flapping, 
\cite{garstang2022instrumented}
flapping being characterized by oscillations with a particular amplitude and frequency 
(see Materials and Methods). 
We focus on the interval of $\tau$ between about 0.1 and 10\,s, 
corresponding to length scales $\ell = \tau V$ between 1 and 100\,m 
at a typical flight speed, $V$, of approximately 10\,m/s, 
since it is in this interval of the spectrum of accelerations that turbulence dominates over other behaviors and interactions. 
\cite{laurent2021turbulence}

For straight and level flight, 
the eagles' accelerations are equal to the acceleration of gravity 
and the acceleration differences are equal to zero. 
Changes in the lift force produce vertical accelerations 
that we detected 
up to small corrections for fluctuations in the eagles' pitch and roll angles. 
We argue that small fluctuations in the vertical wind velocity, 
$w_z$, 
generated small changes in the lift force, 
so that $a_z = w_z / \tau_b$ at leading order, 
where $\tau_b$ is a characteristic timescale of the eagle (about 1\,s). 
\cite{laurent2021turbulence}
This linear relation is sufficient to explain the spectrum of accelerations 
over a broad range of wind conditions 
(Supplementary Materials and Laurent \textit{et al.}~\cite{laurent2021turbulence}). 
In what follows, 
we show that the linear relation does not hold in extreme events, 
and that these extreme events are likely to correspond to wind gusts 
rather than to some other aspect of eagle behavior. 

\begin{figure}[ht] 
 \centering
   \includegraphics[width=1\textwidth, trim={0.5 225 0.5 250},clip]{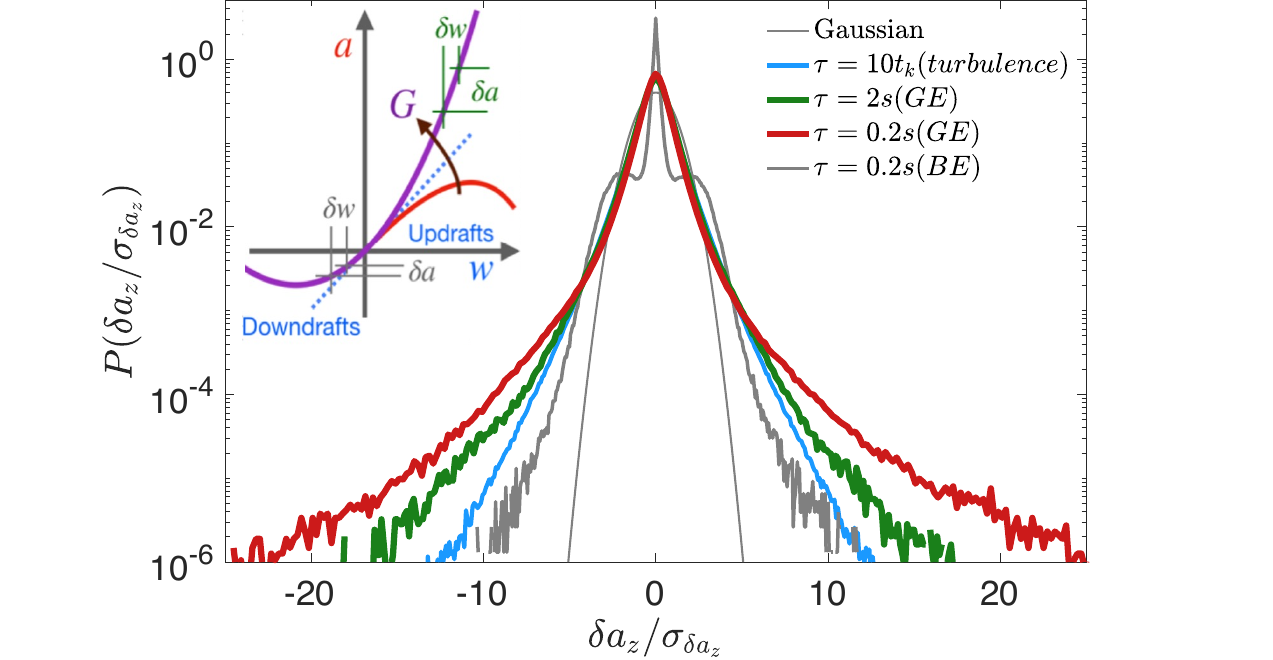}  
    \caption{The probability distributions of eagles' acceleration differences exhibit long tails 
    characteristic of turbulence and of intermittency (blue), and not of gust mitigation, 
    whose purpose is to suppress large accelerations.  
    The tails are broader on short timescales (0.2\,s, red) than long ones (2\,s, green).  
    In gliding or soaring flight (red, green, golden eagles (GE)), 
    the tails exhibit no preferred value of $\delta a_z$ 
    and decay monotonically toward large $|\delta a_z|$ in the same way as turbulence.  
    The distributions of accelerations throughout flight (grey, bald eagles (BE)), on the other hand, 
    exhibit pronounced shoulders at the values of $\delta a_z$ characteristic of flapping.  
    The inset illustrates the nonlinear model we propose (purple) 
    that explains the relationship between turbulent wind speed fluctuations, $w$, 
    and the accelerations we measured, $a$.  
    The linear model (blue dotted) explains the acceleration spectrum but not extreme events, 
    and gust mitigation or aerodynamic stall generate curves that bend down (red) rather than up. Note that $t_k$ represents the time required for fluid particles to traverse a turbulent eddy of the Kolmogorov length scale.}
    
   \label{fig:figure1}
\end{figure}

\subsection*{Distribution of the eagles' acceleration differences}
As seen in Fig.~\ref{fig:figure1}, 
the acceleration differences for soaring golden eagles are intermittent 
in the sense that their probability distributions have broad tails. 
The broad tails reflect the occurrence of extreme events 
as strong as 25 times the standard deviation, 
which would not have been observable had the data followed Gaussian statistics. 
Furthermore, the accelerations are more intermittent on short timescales (0.2\,s) 
than long ones (2\,s), 
consistent with the increasing intermittency of turbulence toward small scales. 
The distributions are similar to turbulence in that they lack a preferred scale. 
To illustrate this point, we compare the accelerations of golden eagles while gliding 
with those of bald eagles throughout flight. 
In contrast to the golden eagle distributions, 
the bald eagle distributions exhibit shoulders 
centered on particular values of the acceleration that characterize flapping. 
We next address the question of what might be responsible for 
the extreme events during gliding flight. 

\subsection*{Vortical structure of the eagles’ acceleration}
The dynamics of the accelerations in the vicinity of extreme events 
are consistent with encounters with turbulent vortices. 
To show this, we looked in the vicinity of large excursions in the acceleration differences, 
and calculated the average of all $\delta a_z(t-t_0)$ and $a_z(t-t_0)$ 
for which 
$|\delta a_z(t_0)| > H \sigma_{\delta a_z}$, 
where $t_0$ are the times of local extrema in $\delta a_z(t)$, 
$H$ is a multiple of the standard deviation of the relative accelerations, 
$\sigma_{\delta a_z} \equiv \langle \delta a_z^2 \rangle^{1/2}$, 
and the brackets denote averaging over all $t$. 
For linear relationships between wind velocity fluctuations and eagle accelerations, 
the statistics of the acceleration differences are the same as those of the wind velocity fluctuations, 
since $\delta a_z = \delta w_z / \tau_b$, 
where $\delta w_z \equiv w_z(t+\tau/2) - w_z(t-\tau/2)$. 
We compare the eagle acceleration statistics 
with the Eulerian statistics of transverse velocity increments, $\delta w_z(\ell)$, 
a comparison that is quantitative for approximately straight flight trajectories 
and flight speeds that are large relative to the wind velocity fluctuations. 
\cite{taylor1938spectrum, laurent2021turbulence}
The wind velocity statistics come from a wind tunnel experiment 
(Materials and Methods). 

\begin{figure}[ht]  
   \centering
   \includegraphics[width=1\textwidth, trim={05 400 5 80},clip]{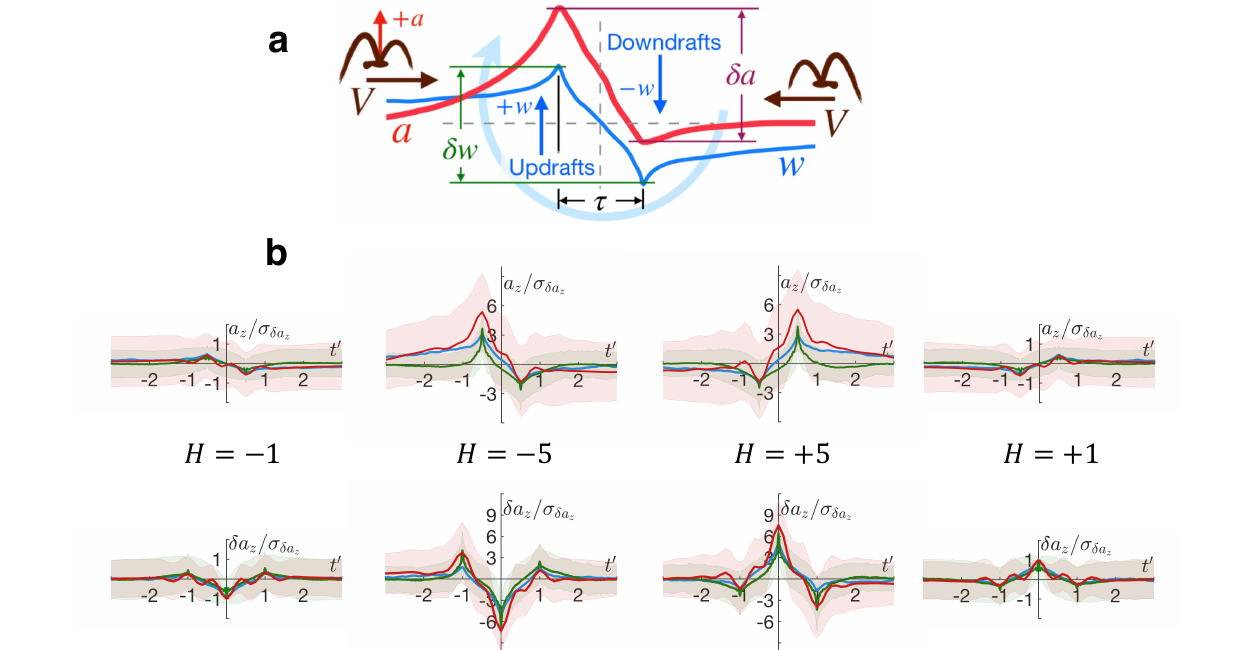}  
   \caption{The evolution of eagle accelerations during extreme events 
   is consistent with gust encounters. 
   (a) When flying through turbulence, 
   eagles generically encounter vortical winds ($w$, blue), 
   that push them up and then down (or vice versa).  
   By Newton’s law, we relate the forces produced by these winds with the eagles’ accelerations 
   ($a = F/m$, red).  
   The velocity and acceleration differences, made across times $\tau$, 
   isolate vortices of size $\ell = \tau V$ from the continuum produced by turbulence, 
   and the possible effects of vortices on eagles.  
   (b) Near local extrema in the acceleration differences, 
   averaged eagle accelerations (above) 
   and their differences (below) 
   resemble averaged wind velocities and their differences (blue). 
   On the far left and far right ($H = \pm 1$), 
   the winds and accelerations have the same symmetries.  
   For extreme events ($H = \pm 5$, center left and center right), however, 
   the eagles accelerated upward more than downward, 
   and more so on small scales ($\tau$ = 0.2\,s, red) 
   than on large scales ($\tau$ = 2\,s, green).  
   The shaded regions indicate $\pm$1 standard deviations of all events that composed the averages 
   ($N \approx$ 10$^6$ for $H = \pm 1$ and $N \approx$ 10$^3$ for $H = \pm 5$)}. 
   \label{fig:figure2}
\end{figure}

Figure \ref{fig:figure2} shows that in any event, large or small, 
eagles were accelerated up and down as is expected from flying through vortices 
that first pushed them up and then down (or vice versa). 
As a consequence of the approximate symmetry in the up- and downward motions of the wind 
(as much air goes up as comes down), 
the wind velocity differences and acceleration differences for events of typical amplitude 
($H = \pm 1$) are left-right symmetric about $t_0$, 
and are flipped about $\delta a_z = 0$ with changes in the sign of $H$. 
The corresponding wind velocities and eagle accelerations are mirror-symmetric, 
the former describing the profile of the vertical wind velocities encountered 
while traversing a vortex, 
and the latter the corresponding changes in the force on the eagles. 
These findings hold for any $\tau$ in the interval between about 0.2 and 4\,s, 
corresponding to vortices with length scales, $\ell$, of between 2 and 40\,m. 
Note that the lower bound is approximately the wingspan of eagles ($b \approx$ 2\,m), 
and that the upper bound approaches the size of the typical large eddy in the atmosphere 
(about 100\,m). 
Since the interactions observed in Fig.~\ref{fig:figure2} endure for at least $2 \tau$, 
we can imagine the eagles traversing structures that are coherent over scales of between 4 and 80\,m. 
We note an oscillation with a short timescale 
that is more pronounced in the eagle accelerations than in the wind velocity data, 
which has its own timescale, and whose origin is unknown to us. 
This oscillation could be related to the way the sensors were attached to the eagles 
or to wing adjustments made in response to turbulent gusts, including partial flaps or tucks. 
\cite{reynolds2014wing}

Figure \ref{fig:figure2} shows that for extreme events ($H = \pm 5$), 
the time-reversal symmetry of the acceleration differences is broken, 
though it is maintained in the wind velocity differences. 
This broken symmetry is a consequence of accelerations that predominately pushed the eagles upward, 
with downward accelerations being relatively weak. 
This bias in favor of upward accelerations and increased lift is in the eagles' interest to stay aloft, 
and is not consistent with linearity in their response to randomly sampled winds. 
Turbulence itself is generally anisotropic at large scales in the atmosphere, 
but this anisotropy diminishes toward small scales 
where transverse velocity statistics are increasingly symmetric. 
\cite{kurien2000anisotropic, biferale2005anisotropy}
This increasing symmetry in turbulence toward small scales 
is the opposite of the increasing asymmetry we observe in the eagles' accelerations. 

\subsection*{Nonlinear model of the eagles' response to gusts}
The broken symmetry in the eagles' accelerations 
indicates a nonlinear response to the turbulent fluctuations, 
as illustrated in Fig.~\ref{fig:figure1}, \textit{inset}. 
Motivated by the observation that the nonlinearity appears for strong and short gusts, 
we introduce a correction to the linear model of the form 
$a_z = (w_z/\tau_b)(1 + k w_z \tau_b/\ell)$, 
where $k$ is an unknown and dimensionless coefficient that we can estimate from the data. 
For $k<0$, the correction describes a decrease of lift in strong gusts relative to the linear model, 
consistent with aerodynamic stall or gust mitigation strategies. 
For $k>0$, on the other hand, 
strong gusts generate anomalously large vertical accelerations 
and excess lift consistent with gust amplification.

To estimate the strength and sign of the nonlinearity, $k$, 
in the eagles' responses to strong fluctuations, 
we note that gusts tend to blow upward, by $w_z^+$, say, 
as much as downward, so that $w_z^- = -w_z^+$. 
At leading order in the gust acceleration, 
$G \equiv w_z \tau_b / \ell$, 
and assuming isotropic small-scale wind velocity differences, 
we find that $a_z^+/a_z^- \approx -1 + k (\tau_b^2 / \ell)(a_z^+ - a_z^-)$, 
where $a_z^+$ and $a_z^-$ are the peak up- and downward accelerations 
we observed in the eagles during extreme events. 
In other words, we can estimate $k$ from the asymmetry in the excursions in the eagles' accelerations. 
We find that $k = 0.6 \pm 0.1$ in the interval 0.2\,s $< \tau <$ 2\,s, 
using the values for $V$ and $\tau_b$ given above. 
Consistent with the notion of gust amplification, we find that $k > 0$. 

The nonlinear model predicts the observation that short gusts 
are more intermittent than long ones in a manner consistent with, 
but more dramatic than, 
the increasing intermittency of turbulence toward small scales. 
The long tails in the acceleration difference distributions (Fig.~\ref{fig:figure1}) 
suggest that strong accelerations impacted the eagles 
with even higher probability than turbulence predicts according to a linear model. 
One way to quantify intermittency is with the flatness, 
which describes the breadth of the tails of a distribution, 
and which is larger for more intermittent quantities. 
According to a linear model, 
the flatness of the acceleration differences scales in the same way as the difference in wind speed, 
$F_{\delta a_z} \sim F_{\delta w_z}$, 
where $F_{\delta q} \equiv \langle \delta q^4 \rangle / \langle \delta q^2 \rangle^2$, 
with $q$ being either $a_z$ or $w_z$. 
According to the nonlinear model, 
we find that $F_{\delta a_z} \approx (1 + 4 k \sigma_{w_z} \tau_b / \ell) F_{\delta w_z}$ 
at leading order in the gust acceleration, 
with $\sigma_{w_z} \equiv \langle w_z^2 \rangle^{1/2}$, 
and where we calculated the flatness only for positive excursions in $w_z$. 
The latter is necessary since the nonlinearity is not physical 
for large negative fluctuations in $w_z$, 
and is justified since only the positive excursions contribute to large $|\delta a_z|$ 
and to $F_{\delta a_z}$ in our model. 
Note that the acceleration difference distribution is approximately symmetric in the nonlinear model, 
even if the accelerations themselves are asymmetric, 
since large values of $\pm\delta a_z$ will be encountered on both the ascending and descending limbs of large excursions in $a_z(w_z)$ 
(see Fig.~\ref{fig:figure1}, \textit{inset}). 

\begin{figure}[ht]  
\centering
   \includegraphics[width=1\textwidth,trim={0.5 60 0.5 20},clip]{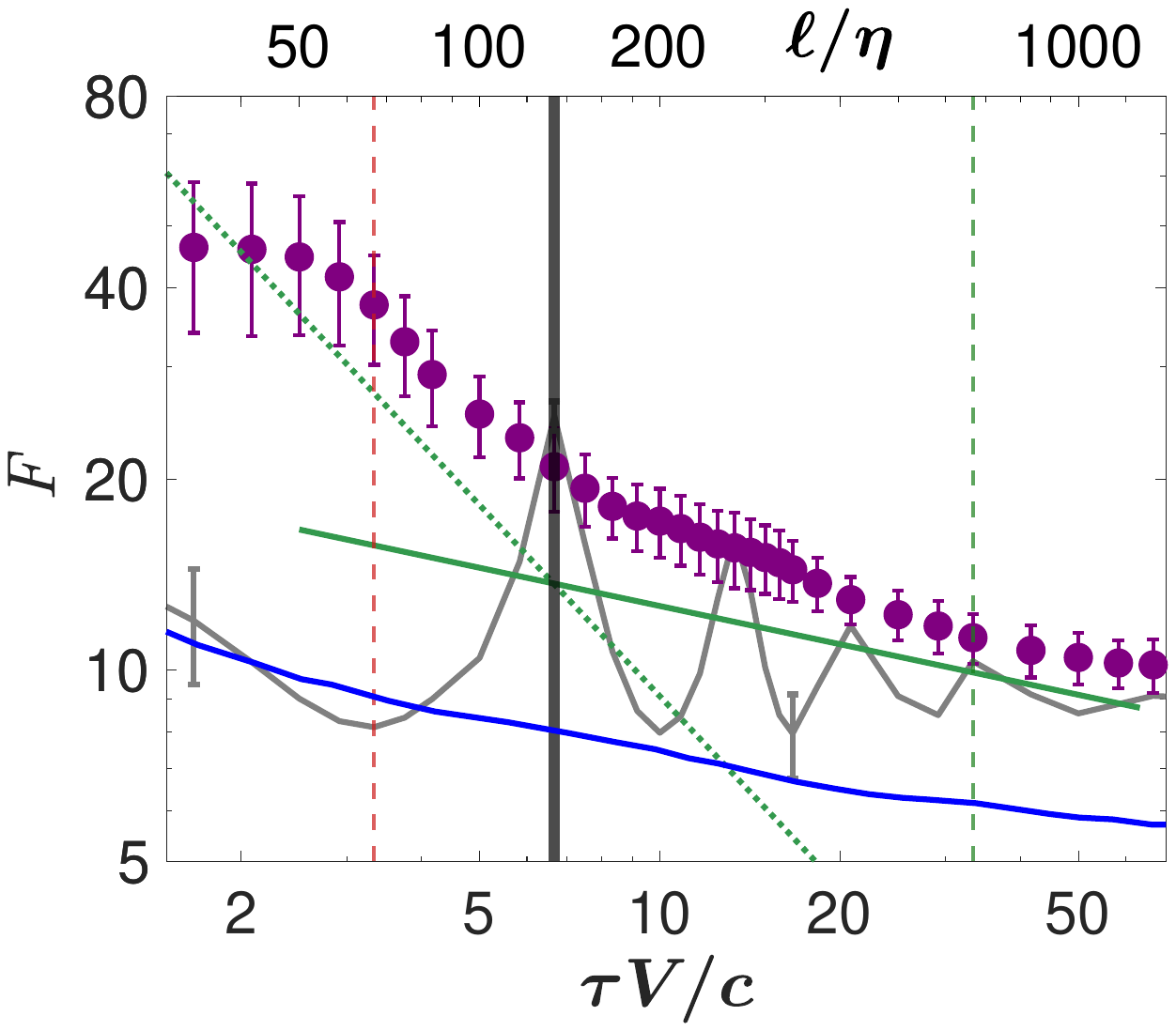}  
   \caption{During gliding flight, 
   extreme fluctuations increased in frequency toward small timescales, $\tau$.  
   For $\tau V/c > 7$, 
   the flatness of acceleration differences ($F_{\delta a_z}$, purple circles) 
   is consistent with the scaling law observed in turbulent wind fluctuations 
   ($\sim\tau^{-0.2}$, solid green line, 
   and $F_{\delta w_z}$, blue curve. \cite{dhruva2000experimental})  
   For $\tau V/c < 7$, $F_{\delta a_z}$ blows up toward small $\tau$ 
   in a way consistent with our nonlinear model ($\sim\tau^{-1}$, dotted green line).  Note that \textit{c} = 0.6 m is the wing chord length.
   For comparison, the flatness of Gaussian distributions is equal to 3, 
   and for the flight of bald eagles that is dominated by flapping (gray curve), 
   the data exhibit peaks at the characteristic timescales of flapping.  
   The error bars are the standard error. 
   The vertical lines (dashed red and green) 
   indicate the timescales for the distributions displayed in Fig.~\ref{fig:figure1}. 
   The upper x-axis applies only for the turbulence data (blue curve), 
   where $\eta$ is the Kolmogorov scale. \cite{kolmogorov1991local}}
   \label{fig:figure3}
\end{figure}

Figure \ref{fig:figure3} 
shows that the eagles appear to exhibit the subtle intermittency of turbulence for $\tau >$ 0.4\,s, 
which manifests as a slow rise in $F_{\delta a_z}$ toward small $\tau$ 
consistent with the approximate $\tau^{-0.2}$ scaling observed for turbulent wind speed increments 
on timescales of a few seconds down to a few milliseconds. 
\cite{dhruva1997transverse}
On shorter timescales for the eagles, $\tau <$ 0.4\,s, however, 
we observe a sharp increase in acceleration intermittency, 
corresponding to the increase in the frequency of extreme events seen in Fig.~\ref{fig:figure1}. 
This increase is incompatible with either a linear response to turbulence, 
which would yield an extension of the $\tau^{-0.2}$ scaling toward smaller $\tau$, 
or with mechanisms to mitigate or avoid gusts, 
which would attenuate accelerations and decrease acceleration intermittency at any $\tau$.  

The increase in acceleration difference flatness 
observed in Fig.~\ref{fig:figure3} at small scales 
is a feature of the nonlinear model described above. 
The nonlinear model exhibits a crossover near 
$\tau_c = \ell_c / V = 1/(4 V k \sigma_{w_z} \tau_b)$ 
between a large-$\ell$ regime for which the acceleration difference flatness 
is approximately equal to the wind velocity difference flatness, 
and a small-$\ell$ regime 
where the acceleration difference flatness varies predominately as $\ell \sim (\tau V)^{-1}$, 
the latter scaling law holding approximately 
if the dependence of the wind velocity difference flatness on $\tau$ is weak, 
as it is in practice. 
\cite{sreenivasan1997phenomenology}
For the values of $V$, $\tau_b$, and $k$ given previously, 
our prediction is that 
$\tau_c \approx$ 0.3\,s. 
This value compares favorably 
with the crossover observed 
in Fig.~\ref{fig:figure3} near $\tau \approx$ 0.4\,s. 
The sharp increase in the acceleration flatness appears to plateau for small $\tau V / c \ll 1$, 
or for eddies much smaller than an eagle's wing chord. 
The plateau is consistent with a picture of gusts smaller than an eagles' wing chord 
acting non-uniformly over the surface area of the wings 
and being automatically mitigated by averaging over the wing area. 

It is worth noting that the observed crossover timescale between the linear and nonlinear regimes 
is longer 
than the minimum reaction time observed in other species of birds 
like the starling and barn owl ($\approx$80\,ms), 
\cite{pomeroy1977laboratory,cheney2020bird} 
which permits the idea of active engagement by the eagle with turbulence 
through deliberate changes in its attitude or geometry. 
These responses could include changes in 
the wing shape, body angle, angle of attack, or tail angles 
or combination of all as observed in several species. 
\cite{cheney2020bird, usherwood2020high, swiney2020preliminary, tucker1993gliding, norberg2012vertebrate}

\subsection*{Discussion and conclusions}
We briefly discuss the possible impact of the extreme events on the aerodynamics of eagles. 
Assuming that all extreme events observed in the interval 0.2\,s $< \tau <$ 2\,s 
are due to interactions with gusts, 
which is what the statistics suggest, 
the distributions in Fig.~\ref{fig:figure1} 
quantify the probability of occurrence of gusts of any intensity. 
To do this in a physically meaningful way, we estimate the gust ratio, 
$GR \equiv |\delta w_z|/V$, 
or the ratio of vertical wind speed fluctuations to the flight speed. 
Since the dynamics are approximately linear up until gust encounters, we assume linearity to estimate the flight time spent outside of gusts. 
Thus, the gust ratio is $GR \approx |\delta a_z| \tau_b / V$ 
and exceeds a certain value, $B$, when $|\delta a_z| > B V / \tau_b$. 
The probability of these events occurring can be found 
by integration of the tails of the distributions in Fig.~\ref{fig:figure1}, 
noting that $V / \tau_b$ is about 10\,m/s$^2$ 
and approximately equal to the acceleration of gravity. 
The gust ratio required to stall a wing in steady state, 
which we take to happen for angle of attack increases of about 15 degrees, is about 0.3. 
Due to nonlinear aerodynamic effects and the eagles' own reactions, 
we think it is unlikely that the eagles' wings stall, 
but we use this value as an indicator of when strong changes in aerodynamics are likely to happen. 
We find that about 20\% of the flight time was spent in these extreme events ($GR > 0.3$), 
which is not insignificant. 
It appears to us that gusts are not rare perturbations, 
and that soaring flight for golden eagles, as for flapping flight, 
is substantially unsteady and nonlinear.

In summary, we observed that golden eagles in soaring flight experienced short, 
sharp shocks that preferentially pushed the eagles upward. 
To understand this in terms of the eagles' responses to gusts, 
we introduced an empirical definition of gusts that respects the intermittent 
and necessarily statistical description of turbulence. 
In this picture, gusts are vortical structures arising on a continuum of spatiotemporal scales, 
and with amplitudes drawn from a distribution that we measured (Fig.~\ref{fig:figure1}). 
We find the fingerprint of turbulence intermittency in the statistics of the eagle accelerations, 
with extreme events becoming stronger toward small scales 
-- the smaller the size of the gust, the stronger is the gust. 
The extreme events observed in eagles have the same temporal evolution 
as predicted by a simple nonlinear model of their interaction with turbulent gusts 
(Fig.~\ref{fig:figure2}). 
At the smallest scales, 
we observed eagle accelerations to be more intermittent than turbulence (Fig.~\ref{fig:figure3}), 
with strong upward accelerations being favored over downward ones. 

The evidence suggests that gusts are prevalent throughout an eagle's airborne life, 
occurring in one out of every five seconds of flight, 
and that gusts provoke a strongly nonlinear response. 
The response suggests that the eagles ratchet upward on turbulent gusts. 
These dynamics have, to the best of our knowledge, not been observed previously in any animal movements, 
and highlight the potentially beneficial aspects of ubiquitous environmental turbulence 
and its intrinsic intermittency, or gustiness. 
Next steps include acquiring the additional data likely required to understand 
the physical mechanisms underlying the eagles' asymmetric responses. 
We note one possible detrimental consequence of the broad acceleration distributions we observed, 
which is that accelerations strong enough to damage animals 
are much more likely than predicted by Gaussian models. 
This idea suggests that the tails of the distributions must be truncated at some value, 
which we did not observe in our data. 
Smaller birds and insects \cite{combes2009turbulence, shepard2016moving} 
are likely even more susceptible to turbulence intermittency than the eagles we studied, 
and we anticipate work toward understanding the interactions with turbulence of other species 
as well as of our engineered vehicles. 

\bibliography{sn-bibliography}

\newpage
\section*{Materials and Methods}

\subsection*{Data acquisition}
\subsubsection*{Eagle Flight Acceleration data}

We observed six golden eagles (\textit{A. chrysaetos}) and six bald eagles (\textit{H. leucocephalus}). Golden eagles weighed between 3760 – 6220 g with wing chords from 569 – 640 mm, and bald eagles weighed from 3950 – 4900 g with wing chords from 590 – 625 mm. An on-board instrumentation package was mounted dorsally 
on each bird using Teflon ribbon (Bally Ribbon Mills, Bally, PA, USA). 
The instrumentation consisted of a triaxial accelerometer 
and global positioning system (GPS) (Cellular Tracking Technology (CTT), LLC, Rio Grande, NJ, USA, weight=70\,g). 
The accelerometer gave measurements of the acceleration vector 
defined in a right-handed coordinate system 
with x-axis aligned longitudinally, y-axis laterally and z-axis vertically. 
We obtained the position and altitude above mean sea level of the eagles from the GPS. 
Eagles flew for an average 26 hours (ranging from 5 to 47 hours) 
for about 17 days from March 10, 2016, to April 2, 2016, 
resulting in a total of 315 hours of flight acceleration data. 
Their flap time varied between 2\% to 38\% of total flight time 
and their flapping frequency from 2.2\,Hz to 2.8\,Hz. 
The accelerometer and the GPS were operating at 40\,Hz and 0.0167\,Hz, respectively. The units were programmed to collect data in bursts at irregular time intervals, 
and to transmit them daily to CTT servers via the global system for mobile communications.    \\ 

\subsubsection*{Wind Tunnel Turbulence data}
To obtain turbulent wind velocity data, 
we carried out experiments in Warhaft Turbulence Wind Tunnel at Cornell University.\cite{yoon1990evolution} 
This wind tunnel has a 0.91\,m by 0.91\,m cross-section, 
a 9.1\,m long test section, 
and can reach mean winds of up to 20\,m/s. 
We generated turbulence using a biplanar square grid 
with a mesh spacing of 0.083\,m at the inlet of the test section, 
and to generate shear 
we covered the upper half of the grid with stainless steel wire cloth of 0.14\,mm opening size, 
thereby giving rise to high-speed flow in the bottom half of the wind tunnel 
and low-speed in its top half. 
The high- and the low-speed flows mix as they evolve downstream, 
giving rising to a shear layer.\cite{gupta2023experimental}

We probed the flow using an x-wire anemometer 
mounted on a 3-axis stepper motor-controlled traverse system. 
The x-wires were 5\,$\mu$m in diameter, 
about 1\,mm long (active sensing part) and 1\,mm apart. 
They were calibrated in the potential core of a jet. 
We took measurements on wind tunnel centreline 7\,m downstream the grid 
at a sampling frequency of 50\,kHz, 
and ensured that the data were long enough for statistical convergence 
of the 4th order moment of velocity increments. 
The Reynolds number, $\sigma_u\lambda/\nu$, was about 700 
where $\sigma_u =\langle u^2 \rangle^{1/2}$ 
is the RMS of the longitudinal turbulence fluctuations 
and $\lambda =\sigma_u/\langle(\partial \sigma_u/\partial x)^2\rangle^{1/2}$ is the Taylor microscale.

\subsection*{Classification of flight behaviors}
The flight data were acquired over periods long enough 
that eagles flapped 
and performed non-flying behaviors like take-off, landing or perching. 
Because flapping presents a dominant and strong contribution 
to the eagles' accelerations, 
we separated flapping from non-flapping flight. 
Furthermore, we removed non-flying behaviors. 
We developed a method to remove the flapping parts of flights 
as described here, 
with special attention to the differences 
from the method described in Laurent {\it et al.} \cite{laurent2021turbulence} 
A distinct signature of the flapping in the data 
is the presence of large oscillation in both x- and z-accelerations. 
We focused first on the z-accelerations to identify the flapping frequency 
since it is affected strongly by the upward and downward motions 
of the wings.\cite{reynolds2014wing}
The Fast Fourier Transform (FFT) on the z-acceleration of each eagle 
reveals a distinct spike at frequencies that varied between 2.2\,Hz and 2.8\,Hz across the eagles, 
which gave us an approximate flapping frequency for each bird. 
The dominant spike in the FFT of acceleration data of each eagle 
was associated with another spike at double the frequency, a harmonic. 
To identify the specific time evolution of the flaps, 
we first identified 
200 flaps by visual inspection of the x- and z-accelerations. 
For this purpose, 
we considered only those flaps 
that were in groups of several unambiguous oscillations 
and that were distinct from neighboring accelerations in magnitude. 
We aligned these manually identified flaps with respect to the peak in the z-accelerations 
and took a mean of the aligned flaps (Fig.~S1). 
The peak of the mean flap occurred at approximately $T/4$, 
where $T$ is the flapping period. 
We labeled the time corresponding to the peak of mean flap as $T$=0 
and used this mean flap to automatically search for flaps 
in the raw acceleration data 
as described next. 

To search for flaps in the x- and z-acceleration automatically, 
we identified all local maximum z-accelerations above or equal to 
a certain value.  
We found that our results were insensitive to 
the particular threshold in the peak value when it was 
near -0.5\,g (Fig.~S2). 
To 
minimize the noise sensitivity of the local maxima, 
we fixed a finite-sized time interval corresponding to each maximum ($\pm 1/4^{th}$ of the flap time) 
and chose as the flap amplitude only the single acceleration 
that was maximum in that interval, 
and labeled its corresponding time as $T_0$. 
We then selected z-accelerations within $-T_{flap/4}$ 
and $3T_{flap}/4$ of $T_0$ 
($T_{flap}$ is flapping time period) 
for each peak and aligned these peaks with the peak of the mean flap (Fig.~S1). 
To determine whether the local maxima were flaps 
and not some other type of event, 
we found the cross-correlation coefficient ($r_z$) 
between the z-accelerations ($a_z$) 
and the mean flap ($\overline {a_z}$ ), 
defined as 
\begin{equation}
 r_z=\frac{1}{(n-1)} \frac{\sum \overline{a_z}a_z }{ \sigma^2_{\overline{a_z}}} 
\end{equation}
where $n$ is the number of data points in each flap. 
Since flaps produce distinctive accelerations 
in both the x- and z-directions, 
we also computed the correlation coefficient ($r_x$) between x-acceleration 
and the mean flap acceleration in x-direction using same method as above. 
We identified only those acceleration intervals as flapping for which both 
$r_z \ge 0.5$ and $r_x \ge 0.5$ (Fig.~S3). To confirm that we removed flapping from the accelerometer data, 
we found that the spike corresponding to the flapping frequency 
was absent from the acceleration spectrum
(Fig.~S4). 

To remove non-flying behaviors like take-off, landing, and perching, 
we adopted the method described in Wilson {\it et al.}\cite{wilson2008prying} 
and Laurent {\it et al.}\cite{laurent2021turbulence}
We first smoothed the x-accelerations using a running mean of 5\,s. 
We then calculated the body pitch as the arcsine of the x-accelerations 
and labeled any accelerations that produced pitch angle larger than 25$^\circ$ 
as non-flying behaviors. 
We also used different lengths for the running mean and pitch angle thresholds 
in order to test for the possible influence of missed non-flying behavior on acceleration, 
and found that our conclusions were not affected for 
running means with lengths in the vicinity of 5\,s (Fig.~S5a) 
and pitch angle thresholds near 25$^\circ$ (Fig.~S5b), 
consistent with Laurent {\it et al.} \cite{laurent2021turbulence}

\subsection*{Conditional Averaging}
In this section, 
we discuss how we extracted the vortical structure of the response of the eagles to gust from the data using z-component of acceleration obtained by the accelerometers. 
In the case of turbulence experiments, several algorithms have been used to detect the extreme events and their associated vortical structures \cite{mouri2003vortex, camussi1999experimental}. One such algorithm is based on velocity differences, 
\textit{i.e.}, once velocity data are acquired by hot-wire anemometers, the difference between instantaneous velocity separated by some time delay is compared with a certain threshold. 
All these events that exceed the given threshold are then aligned and are averaged to obtain a structure that may resemble a cross-section of the vortex. 

We followed a similar technique to study the vortical structure of the eagles’ acceleration. First, we found local maxima in $\delta a (t,\tau)$ that exceed a threshold. Here, $\delta a (t,\tau)$  is instantaneous acceleration difference over time delay, $\tau$, and the threshold is $|H|\delta a_{RMS},$ where $H$ is an integer and $\delta a_{RMS}= <\delta a^2 >^{1/2}$. We labelled the time corresponding to each local maximum as $T$, and selected $\delta (t)$ within $\pm 5T$. Like flap detection algorithm, we then align these $\delta a(t)$ with respect to the peak and averaged them. The numerical factor, H is so chosen that it would correspond to desired magnitude of gust ratio. For lower gust ratio (=0.3), $H=1$, called here as typical events, and $H\ge5$ corresponds to $GR\ge 1,$ which we call as extreme events. It is also essential to note the larger the factor $H$ is, the smaller is the total number of events detected, as can also be seen from Fig.~1, and more significant the statistical uncertainty is. Here, `+' sign corresponds to the response to the updraft event while `--' to the downdraft. The sign of $H$ should not be undermined since it is this difference that allowed us to make a fine distinction in the eagles’ response to updraft and downdraft.

\subsection*{Estimation of time spent in gusts}
We estimated the time spent by eagles in gusts exceeding GR$\approx 0.3$ by integrating the tail of the distribution exceeding  $|\delta a_z|/ \sigma_{\delta a_z} \ge 1$ (Fig. 1), i.e.,
\begin{equation}
P_{GR\ge0.3}= \int_{-\infty}^{-1} P \left (\frac{\delta a_z} {\sigma_{\delta a_z}}\right) \,d\left (\frac{\delta a_z } {\sigma_{\delta a_z}}\right) + \int_{1}^{\infty} P \left (\frac{\delta a_z} {\sigma_{\delta a_z}}\right) \,d\left (\frac{\delta a_z} {\sigma_{\delta a_z}}\right) \newline
\end{equation}
Alternatively, $P_{GR\ge0.3}$ could also be found as 
\[ P_{GR\ge0.3}=1- \int_{-1}^{1} P \left (\frac{\delta a_z (\tau)} {\sigma_{\delta a_z}}\right) \,d\left (\frac{\delta a_z (\tau)} {\sigma_{\delta a_z}}\right) \]

\subsection*{Theory}
We briefly summarize some of the calculations needed to 
relate the linear and nonlinear models for the eagle acceleration differences 
with the moments of the acceleration differences, 
including the variance, skewness and flatness. 
These acceleration differences are related to wind velocity differences 
according to our nonlinear model as  
$\delta a_z = (\delta w_z/\tau_b)(1 + k \Sigma w_z \tau_b/\ell)$, 
which can be derived from the nonlinear model given in the text 
by noting that $\delta (w_z^2) = \delta w_z \, \Sigma w_z$, 
where $\Sigma w_z(t, \tau) \equiv w_z(t+\tau/2) + w_z(t-\tau/2)$. 
To derive simple forms for higher order moments of the differences, 
such as the skewness and flatness, 
we expanded around small fluctuations 
and retained only the lowest order term in $k$. 
We also assumed that turbulence is locally 
isotropic,\cite{kolmogorov1991local} 
and that velocity sums and differences are uncorrelated \cite{blum2011signatures} 
so that $\langle \delta w \, \Sigma w \rangle = 0$. 
We always used the linear model to calculate 
the acceleration difference variance, 
which is justified since it is a measure of typical events that we observed 
are approximately linear.\cite{laurent2021turbulence} 
We employed the same approximations 
to estimate the strength of the nonlinearity, $k$.


\section*{Data availability} 
The data that support the findings of this study are available from Cellular Tracking Technologies, but restrictions apply to the availability of these data, which were used under license for the current study, and so are not publicly available. Data are, however, available from the authors upon reasonable request and with permission of Cellular Tracking Technologies and Friends of Talladega National Forest.

\section*{Acknowledgements} 
We recognize stimulating comments from C.M. White, V. Trimble, and T.E. Katzner.

\section*{Author contribution}
 G.P.B. supervised the project; G.P.B. designed research;  M.J.L. and T.M. performed research and provided data; D.G. and G.P.B. analyzed data; D.G. and G.P.B. wrote the paper; All authors discussed the results, and D.B., M.J.L., and T.M. revised the manuscript critically for important intellectual content.

\section*{Competing interests}
The authors declare no competing interests.

\section*{Ethics approval} 
Use of golden and bald eagles for this research was approved by the West Virginia University Institutional Animal Care and Use Committee (Protocol 11-0304).


\end{document}